\documentclass[prl,twocolumn,showpacs,amsfonts,amsmath,amssymb]{revtex4-1}%

\usepackage{graphicx}

\begin{document}

\title{Three-dimensional nodal superconducting gap in single crystals Ba(Fe$_{1-x}$Ni$_x$)$_2$As$_2$}

\author{C.~Martin}
\affiliation{Ames Laboratory and Department of Physics \& Astronomy, Iowa State University, Ames, IA 50011}

\author{H.~Kim}
\affiliation{Ames Laboratory and Department of Physics \& Astronomy, Iowa State University, Ames, IA 50011}

\author{R.~T.~Gordon}
\affiliation{Ames Laboratory and Department of Physics \& Astronomy, Iowa State University, Ames, IA 50011}

\author{N.~Ni}
\affiliation{Ames Laboratory and Department of Physics \& Astronomy, Iowa State University, Ames, IA 50011}

\author{V.~G.~Kogan}
\affiliation{Ames Laboratory and Department of Physics \& Astronomy, Iowa State University, Ames, IA 50011}

\author{S.~L.~Bud'ko}
\affiliation{Ames Laboratory and Department of Physics \& Astronomy, Iowa State University, Ames, IA 50011}

\author{P.~C.~Canfield}
\affiliation{Ames Laboratory and Department of Physics \& Astronomy, Iowa State University, Ames, IA 50011}

\author{M.~A.~Tanatar}
\affiliation{Ames Laboratory and Department of Physics \& Astronomy, Iowa State University, Ames, IA 50011}
%\affiliation{Ames Laboratory, Iowa State University, Ames, IA 50011}

\author{R.~Prozorov}
\email[Corresponding author: ]{prozorov@ameslab.gov}
\affiliation{Ames Laboratory and Department of Physics \& Astronomy, Iowa State University, Ames, IA 50011}

\date{12 November 2009}

\begin{abstract}
The London penetration depth, $\lambda$, is directly related to the density, $n_{s}$, of the Cooper pairs ($\lambda^{2}\propto 1/n_{s}$) and its variation with temperature provides valuable insight into the pairing mechanism. Here we study the evolution with doping of the temperature dependence of the in-plane ($\lambda_{ab}$) and out-of-plane ($\lambda_{c}$) penetration depths in single crystals of electron-doped Ba(Fe$_{1-x}$Ni$_x$)$_2$As$_2$. As is the case for other pnictides, $\lambda(T) \sim T^n$ over the whole doping range and this behavior extends down to at least $T=T_c/100$, setting a very small upper limit on the gap minimum. Furthermore, in the overdoped regime: 1) the exponent $n$ becomes substantially smaller than 2, which is incompatible with the models that explain power-law behavior to be due to scattering; 2) the exponent $n$ becomes anisotropic, with $\lambda_{c}(T)$ showing a clear $T$-linear behavior over a large temperature interval. These findings suggest that in the overdoped regime the superconducting gap in iron-based pnictide superconductors develops nodal structure, which unlike in the cuprates, cannot be understood within a two-dimensional picture.

\end{abstract}
\maketitle

Determining the symmetry of the superconducting gap function, $\Delta({\bf k})$, is one of the milestones in understanding the pairing mechanisms of superconductors. In iron-based pnictide superconductors the symmetry of the gap is still a debated issue both theoretically and experimentally. Some early measurements, including angle - resolved electron spectroscopy ~\cite{Kaminski08,Ding08,Zhao08,Nakayama08}, point contact spectroscopy~\cite{Szabo08}, penetration depth \cite{carrington,Hashimoto09} and specific heat~\cite{Mu08} were interpreted in a framework of fully gapped superconductivity. Others, such as NMR \cite{Fukazawa09}, penetration depth \cite{Martin1111,MartinPRBBaK,GordonPRB,GordonPRL} and thermal conductivity \cite{Luo09, Machida09} have suggested a strongly anisotropic gap. More recent systematic measurements of thermal conductivity over the whole doping phase diagram in Ba(Fe$_{1-x}$Co$_x$)$_2$As$_2$ (FeCo-122) single crystals have shown a clear evolution towards a more anisotropic gap in the overdoped regime \cite{Tanatarheattransport}. Several other studies of the overdoped compositions have even suggested nodes in the superconduction gap \cite{Fukazawa09,KFe2As2}. Other works have proposed a non-universal gap structure in different pnictide families \cite{Matsuda}. Such a diversity of observations and interpretations is due to several factors, including a substantially three-dimensional character of the Fermi surface and the gap \cite{anisotropy}, competition of superconductivity and other ordered states in the underdoped regime and a possible significant (pair breaking) contribution from scattering. Additional complications are due to complex chemistry and poor control of the composition and crystallinity, especially in RFeAsO (1111) and BaK-122 families of compounds. High quality homogeneous  single crystals are required to address questions related to the gap structure.

In BaFe$_2$As$_2$, superconductivity can be induced by partial substitution of Fe with Co, Pd, Ni, Rh and even combinations of Co and Cu \cite{AthenaCo,NiNiCo,FisherCo,DrNiNi,Canfield09}. Large homogeneous single crystals can be prepared over the whole phase diagram and the doping levels can be controlled and determined with high precision  \cite{NiNiCo,DrNiNi,Canfield09}. In the overdoped regime, where superconductivity is not affected by the coexisting magnetic order (as is the case in the underdoped samples), the superconducting $T_c$ scales well with the number of dopant electrons, obtained by assuming a rigid band approximation \cite{DrNiNi,Canfield09}.

Among the experimental probes of superconductivity, the London penetration depth, $\lambda$, is directly related to the density, $n_{s}$, of the Cooper pairs ($\lambda^{2}\propto 1/n_{s}$), and its variation with temperature provides a valuable insight into the pairing mechanism. For a superconductor with a gap that is nearly uniform in momentum space, $\Delta({\bf k})\approx\Delta_{0}$, the density of quasiparticles, $n_{n}$, is exponentially small at low temperatures ($n_{n}=1-n_{s}$) and $\lambda(T)$ is expected to saturate exponentially below $T\approx T_{c}/3$. On the other hand, in cuprate high temperature superconductors, the penetration depth shows linear temperature dependence ($\lambda\propto T$), consistent with a gap having line nodes~\cite{Hardy93}.

In the present work, we have studied the evolution of both the in-plane ($\lambda_{ab}(T)$) and  inter-plane ($\lambda_{c}(T)$) London penetration depths in single crystals of electron-doped Ba(Fe$_{1-x}$Ni$_{x}$)$_{2}$As$_{2}$ (FeNi-122). We have found that close to the optimal doping level, $x_{opt}$=0.046 \cite{WDS}, $\lambda_{ab}(T)$ is best described by the power-law, $\lambda_{ab}(T)\sim T^n$, with $n\geq 2$, similar to previous results on FeCo-122~\cite{GordonPRL,GordonPRB}. In the overdoped regime however, the exponent $n$ decreases significantly below two and $\lambda_{ab}(T)$ shows a much stronger temperature variation. An even more drastic evolution of $\lambda(T)$ is observed in the inter-plane penetration depth. On the under-doped side, $\lambda_{c}(T)$ shows a tendency to saturation at low temperatures, while in the over-doped regime it is $T-$linear below T$_{c}/3$. These results imply a doping-induced evolution towards a superconducting gap with three-dimensional, nodal structure in the overdoped regime.

Single crystals of electron doped Ba(Fe$_{1-x}$Ni$_{x}$)$_{2}$As$_{2}$ were selected from the growth described in Ref.~\cite{Canfield09}. X-ray diffraction, resistivity, magnetization, magneto-optics and wavelength dispersive spectroscopy elemental analysis (WDS) have all consistently shown good quality single crystals with a small variation of the dopant concentration over the sample and sharp superconducting transitions~\cite{Canfield09}.

The penetration depth was measured using a tunnel diode resonator (TDR) technique~\cite{Degrift74}. The sample was situated inside an inductor with ac magnetic field $H_{ac}\approx 20$\,mOe. Variation of $\lambda(T)$ change the sample's screening ability, and thus the effective inductance and the resonant frequency. The frequency shift, $\Delta f$, is proportional to $\Delta \lambda$. Details of the measurements and data analysis are described elsewhere \cite{Prozorov00}. For $H_{ac}\parallel c$, screening currents flow only in the $ab$-plane and $\Delta f$ is only related to the in-plane penetration depth, $\Delta\lambda_{ab}$, through $\Delta f/\Delta f^{ab}_{0}=\Delta\lambda_{ab}/R_{ab}$, where $\Delta f^{ab}_{0}$ is the total frequency shift when sample is inserted into the resonator and $R_{ab}$ is an effective sample dimension that accounts for the shape and the finite thickness of the sample ~\cite{Prozorov00, Prozorov06}.
When the magnetic field is applied along the $ab$-plane ($H_{ac}\parallel ab$), screening currents flow both in the plane and between the planes, along the $c$-axis. This situation is depicted in Fig.~\ref{Fig3}. In this case, $\Delta f^{\perp}$ contains contributions from both $\Delta\lambda_{ab}$ and $\Delta\lambda_{c}$. For a rectangular sample of thicknesses $2t$, width $2w$ and length $l$, $\Delta f^{\perp}$ is given by Eq.~(\ref{eqmix})

\begin{equation}
\frac{\Delta f^{\perp}}{\Delta f^{\perp}_{0}} \approx \frac{\Delta\lambda_{ab}}{t}+\frac{\Delta\lambda_{c}}{w}=\frac{\Delta\lambda_{mix}}{R_b}
\label{eqmix}
\end{equation}

\noindent where $R_b$ is the effective dimension \cite{Prozorov00} in this configuration and the excitation field $H_{ac}$ is parallel to the longest side $l$ as illustrated on the top sketch in Fig.~\ref{Fig3}. Knowing $\Delta\lambda_{ab}$ from the measurements with $H_{ac}\parallel c$ and the sample dimensions, one can obtain $\Delta\lambda_{c}$ from Eq.~(\ref{eqmix}). However, because $2w\geq 4\times 2t$ in most cases, $\Delta f^{\perp}$ is in general dominated by the contribution from $\Delta\lambda_{ab}$. The subtraction of $\Delta\lambda_{c}$ becomes therefore prone to large errors. The alternative, more accurate approach is to measure the sample twice \cite{cutting}. After the first measurement with the field along the longest side $l$ ($H_{ac}\parallel l$), the sample is cut along this $l$ direction in two halves, so that the width (originally $2w$) is reduced to $w$. Since the thickness $2t$ remains the same, we can now use Eq.~(\ref{eqmix}) to calculate $\Delta\lambda_{c}$ without knowing $\Delta\lambda_{ab}$. Note that the length $l$ and width $w$ are in the crystallographic $ab-$plane, whereas the thickness $2t$ is measured along the $c-$axis. In our experiments, both approaches to estimate $\Delta \lambda_c(T)$ produced the same temperature dependence, but the former technique had a larger data scatter as expected. We therefore only report $\Delta\lambda_{c}$ obtained by cutting and remeasuring the sample.

\begin{figure}[tb]
\begin{center}
\includegraphics[width=9cm]{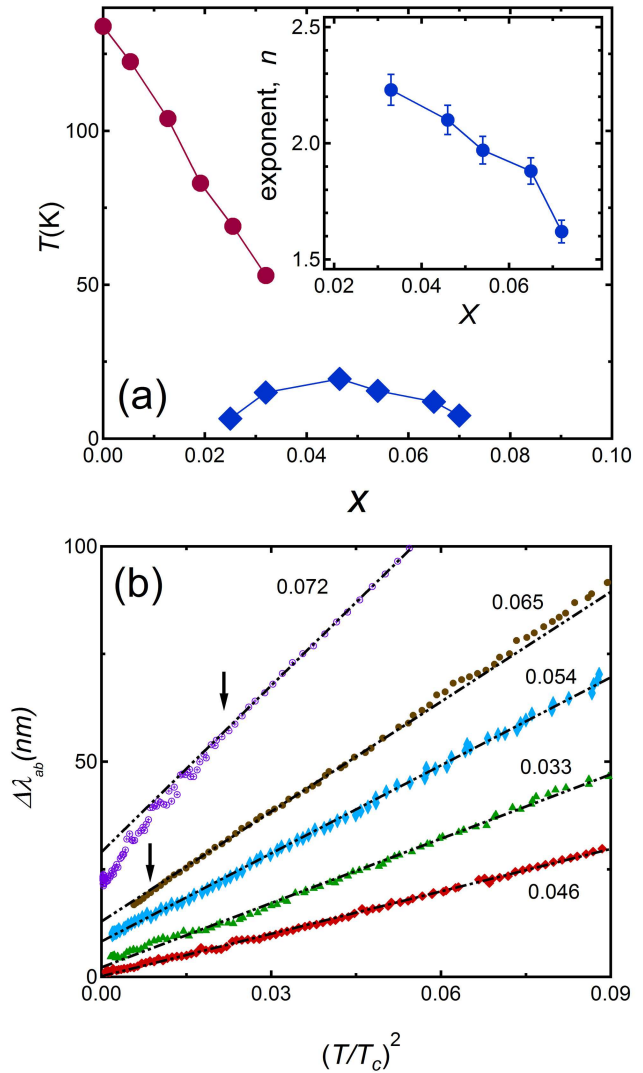}
\end{center}
\caption{(Color online) (a) Temperature-doping phase diagram of Ba(Fe$_{1-x}$Ni$_x$)$_2$As$_2$. The inset shows the power-law exponent $n(x)$ obtained by fitting to $\lambda _{ab} (T)= a+bT^n$ from the base temperature up to $T/T_c$=0.3. (b) $\Delta \lambda _{ab} (T)$ for different doping levels plotted vs. $(T/T_{c})^{2}$. The arrows mark the temperature below which $n$ becomes less than 2. The curves are shifted vertically for clarity.}
\label{Fig1}
\end{figure}

All measured samples showed sharp superconducting transitions with $\Delta T\leq 1K$ (determined as the onset of a diamagnetic signal), consistent with transport and magnetization measurements \cite{Canfield09}. Figure~\ref{Fig1}(a) summarizes the $T(x)$ phase diagram showing structural/magnetic and superconducting transitions. Figure~\ref{Fig1}(b) shows the low-temperature ($T\leq 0.3T_{c}$) behavior of the in-plane penetration depth for several Ni concentrations. The data plotted versus $(T/T_c)^2$ are linear for underdoped compositions and show a clear deviation towards a smaller power-law exponent (below temperatures marked by arrows in Fig.~\ref{Fig1}(b)) for overdoped samples. While at moderate doping levels the results are fully consistent with our previous measurements in FeCo-122 \cite{GordonPRL, GordonPRB}, the behavior in the overdoped samples is clearly less quadratic. It should be noticed that in order to suppress $T_c$ by the same amount, one needs a two times lower Ni concentration compared to Co. In FeCo-122, the samples never reached highly overdoped compositions equivalent to $x=0.072$ of Ni shown in Fig.~\ref{Fig1}(b). Therefore, Ni doping has the advantage of spanning the phase diagram with smaller concentrations of dopant ions, which may act as the scattering centers. The evolution of the exponent $n$ with $x$ is summarized in the inset in Fig.~\ref{Fig1}(a). These values were determined by fitting the curves to $\Delta \lambda_{ab} = a+bT^n$ over the whole temperature range shown in Fig.~\ref{Fig1}(b). If fitted over a temperature range with a smaller upper limit, the overdoped samples show an exponent $n$ significantly smaller than 2.

\begin{figure}%[tb]
\begin{center}
\includegraphics[width=9cm]{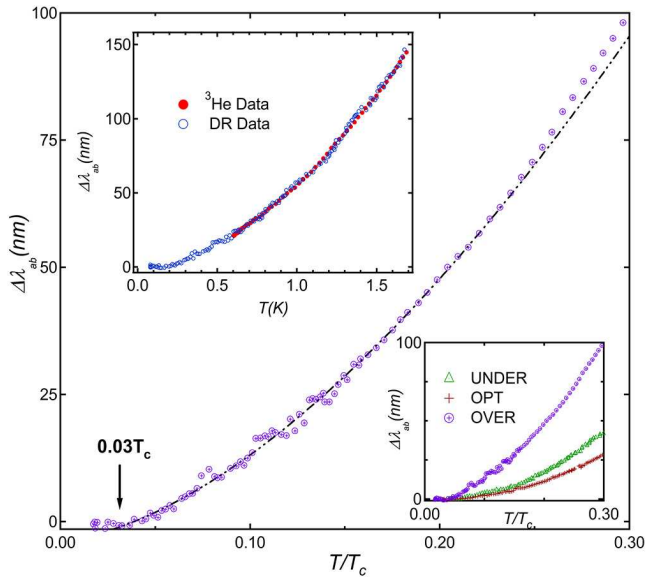}
\end{center}
\caption{(Color online) Main panel: $\Delta \lambda _{ab}(T)$ for an overdoped sample with $x=0.072$, $T_c=7.5$ K, (symbols) and the fit to a power-law
$a+bT^{n}$, $n=1.62$ (dashed line). Upper inset: Data taken with a $^3$He cryostat and with a dilution refrigerator (DR) showing a good overlap. Lower inset: $\Delta \lambda _{ab}(T)$ on the normalized temperature scale comparing  underdoped, $x=0.033$, $T_c=15$ K (UNDER), optimally doped, $x=0.046$, $T_c=19.4$ K (OPT), and overdoped, $x=0.072$, $T_c=7.5$ K (OVER), samples.}
\label{Fig2}
\end{figure}

For any superconducting gap structure, an exponential temperature dependence of $\lambda$ is expected roughly below $T_{min}\approx 0.3 T_c \Delta_{min}/\Delta_{max}$. To estimate $\Delta_{min}$, we used a dilution refrigerator (DR), reaching a base temperature of about 60 mK. As shown in Fig.~\ref{Fig2}, in the overdoped sample the power-law behavior with $n\approx 1.6$ persists at least down to $T/T_c=0.03$, setting an upper limit for the in-plane $\Delta_{min} \approx 0.1 \Delta_{max}$. The upper inset in Fig.~\ref{Fig2} shows a good agreement between DR and $^3$He measurements performed on the same sample. The lower inset compares samples of different doping levels. The slope (parameter $b$ of the fit) correlates with the results found in thermal conductivity measurements \cite{Tanatarheattransport}.

\begin{figure}[tb]
\begin{center}
\includegraphics[width=9cm]{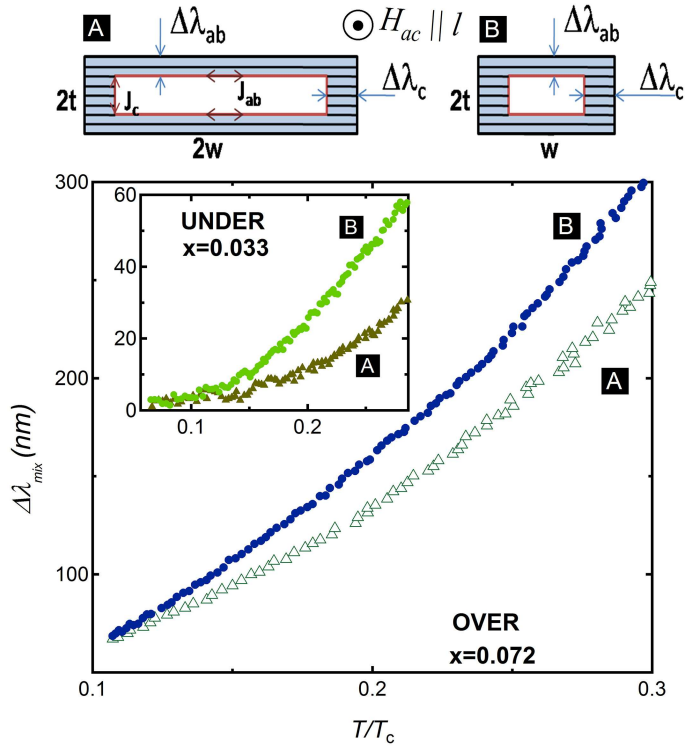}
\end{center}
\caption{(Color online) Schematics of magnetic field penetration in the case of $H_{ac} \parallel l$ for the whole sample [A] and after cutting in half along length $l$ [B]. (Main panel) The mixed penetration depth $\Delta\lambda_{mix}(T)$ before [A] and after [B] cutting for the overdoped sample $x$=0.072, $T_c=7.5$ K. Inset shows similar data for the underdoped $x$=0.033, $T_c=15$ K.}
\label{Fig3}
\end{figure}

\begin{figure}%[tb]
\begin{center}
\includegraphics[width=9cm]{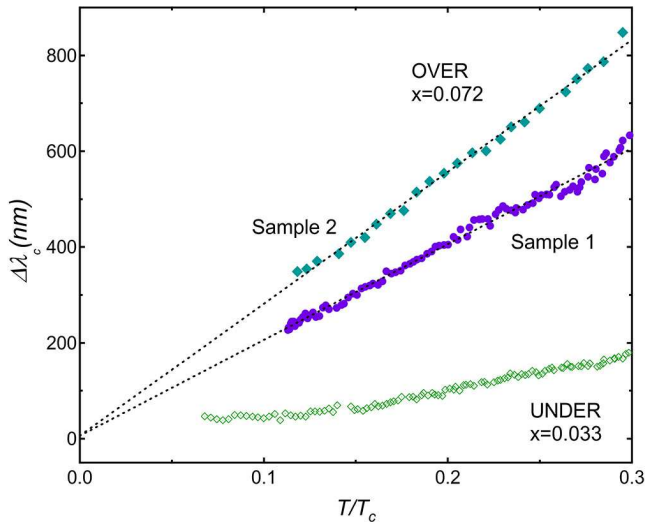}
\end{center}
\caption{(Color online) $\Delta\lambda _{c}(T)$ for the underdoped, $x=0.033$, $T_c=15$ K (UNDER), and for two overdoped, $x=0.072$, $T_c=7.5$ K and $T_c=6.5$ K (OVER), samples. Dashed lines are linear fits.}
\label{Fig4}
\end{figure}

Until now, most studies have focused on the in-plane penetration depth. However, given the three-dimensional character of the band structure, it is imperative to systematically study the inter-plane penetration depth as well. Figure~\ref{Fig3} shows the effective penetration depth, $\lambda_{mix}$ (see Eq.~\ref{eqmix}), for overdoped, $x$=0.072 (main panel), and underdoped, $x$=0.033 (inset), samples before (A) and after (B) cutting in half along the longest side ($l$-side) as illustrated schematically at the top of the figure. Already in the raw data, it is apparent that the overdoped sample exhibits a much smaller exponent $n$ compared to the in-plane penetration depth, while underdoped samples show a tendency to saturate below 0.13T$_{c}$. Using Eq.~\ref{eqmix} we can now extract the true temperature dependent $\Delta \lambda_c(T)$. The result is shown in Fig.~\ref{Fig4} for two different overdoped samples of the same composition, $x=0.072$ having $T_c=7.5$ K and $T_c=6.5$ K, and for an underdoped sample with $x=0.033$ having $T_c=15$ K.
Since the thickness of the sample is smaller than its width, we estimate the resolution of this procedure for $\Delta\lambda _{c}$ to be about 10 nm,
which is much lower than $0.2$ nm for $\Delta\lambda_{ab}$. Nevertheless, the difference between the samples is obvious. The overdoped samples show a clear linear temperature variation up to $T_c/3$, strongly suggesting nodes in the superconducting gap.  The average (between the two samples) slope is large, about $d\lambda_{c}/dT\approx 300 nm/K$ indicating a significant amount of thermally excited quasiparticles. By contrast, in the underdoped sample the inter-plane penetration depth saturates indicating a fully gapped state. If fitted to the power-law the exponent in the underdoped sample $2\leq n\leq 3$, depending on the fitting range.

At face value, our data suggest the development of a three-dimensional nodal structure in the overdoped regime in FeNi-122 crystals. Indeed, the penetration depth depends on the angular average of the response function over the Fermi surface. Nodes, if present somewhere on the Fermi surface, will affect the temperature dependence of both components of $\lambda(T)$. However, the major contribution still comes from the direction of the supercurrent flow, thus placing the nodes in the present case at or close to the poles of the Fermi surface. The nodal topologies that are consistent with our experimental results are latitudinal circular line nodes located at the finite $k_z$ wave vector or a point (or extended area) polar node with a nonlinear ($\Delta(\theta) \sim \theta^p$, $p>1$) variation of the superconducting gap with the polar angle, $\theta$.

Another conclusion from the present study can be made with regard to the importance of scattering. A close to $T^2$ power-law behavior could be explained to be due to scattering in a nodal-gap scenario \cite{Mishra09,hirschfeld}, the proposed extended $s_{\pm}$ symmetry \cite{Mazin08} with a sign change between different sheets of the Fermi surface \cite{Vorontsov09} or strong pair-breaking that drives the system into a gapless regime \cite{Kogan09}. However, neither of these models can explain a close to linear variation of the penetration depth and the anisotropy of the power-law exponent, $n$ (unless scattering is significantly anisotropic). While scattering may still play an important role in determining the exponent $n$, our results suggest a highly anisotropic nodal three-dimensional superconducting gap in the overdoped regime. In other words, not only is the gap not universal across different pnictide families \cite{Matsuda}, it is not universal even withing the same family over an extended doping range.

We thank J.~Schmalian, P.~Hirschfeld, A. Chubukov and I. Mazin for stimulating  discussions. Work at the Ames Laboratory was supported by the Department of Energy  - Office of Basic Energy Sciences under Contract No. DE-AC02-07CH11358. R.P. acknowledges support from the Alfred P. Sloan Foundation.


\begin{thebibliography}{99}
\bibitem{WDS} We use the actual values of $x$ as determined from wavelength dispersive spectroscopy (WDS) microanalysis.
\end{thebibliography}
\end{document}